\documentclass[%
 reprint,
 superscriptaddress,
 amsmath,amssymb,
 aps,
floatfix,
]{revtex4-2}

\usepackage{graphicx}
\usepackage{dcolumn}
\usepackage{bm}
\usepackage{hyperref}
\usepackage[mathlines]{lineno}
\usepackage{float}

\begin{document}

\preprint{APS/123-QED}

\title{Inverse design of plasma metamaterial devices for optical computing}

\author{Jesse A. Rodr\'iguez}
\email{jrodrig@stanford.edu}
\affiliation{Department of Mechanical Engineering, Stanford University, Stanford, CA 94305}
\author{Ahmed I. Abdalla}
\affiliation{Department of Physics, Stanford University, Stanford, CA 94305}
\author{Benjamin Wang}
\affiliation{Department of Mechanical Engineering, Stanford University, Stanford, CA 94305}
\author{Beicheng Lou}
\affiliation{Department of Electrical Engineering, Stanford University, Stanford, CA 94305}
\author{Shanhui Fan}
\affiliation{Department of Electrical Engineering, Stanford University, Stanford, CA 94305}
\author{Mark A. Cappelli}
\affiliation{Department of Mechanical Engineering, Stanford University, Stanford, CA 94305}

\date{\today}

\begin{abstract}
We apply inverse design methods to produce two-dimensional plasma metamaterial (PMM) devices. Backpropagated finite difference frequency domain (FDFD) simulations are used to design waveguides and demultiplexers operating under both transverse electric (TE) and transverse magnetic (TM) modes. Demultiplexing and waveguiding are demonstrated for devices composed of plasma elements with reasonable plasma densities $\sim7$ GHz, allowing for future \textit{in-situ} training and experimental realization of these designs. We also explore the possible applicability of PMMs to nonlinear boolean operations for use in optical computing. Functionally complete logical connectives (OR and AND) are achieved in the TM mode.
\end{abstract}

                    
\maketitle

\section{Introduction}
Inverse design methodology is an algorithmic technique by which optimal solutions are iteratively approximated from an a priori set of performance metrics. When applied to electromagnetically-active systems \cite{minkov2020inverse, ceviche, Su2020Spins, burger2004inverse, BorelTopologyWaveguide, LinPhotonicDirac, miller2013photonic,  ChristiansenTunableLens, PestourieMetasurfacesDemultiplex, ChungTunableMetasurface, MeemInverseFlatLens, piggott2015inversebinarized, liu2013transformation, hughes2018adjoint, molesky2018inverse, christiansen2021inverse,  designsurvey, AndradeInverseGuide}, devices are built by designating a design region wherein various numerical methods are used to solve Maxwell’s equations for a given distribution of sources and material domain. The constitutive properties (e.g., relative permittivity or permeability) of the design region are then mapped to trainable parameters which are modified to mold the propagation of source fields through the designated region in a manner that fulfills the user's performance criteria. Typically, the performance criteria is encoded as an objective function that needs to be maximized or minimized. In many problems, the objective functions involve an inner product between the realized and the desired electromagnetic fields in regions of interest. Given Maxwell’s equations as a constraint, automatic differentiation allows the calculation of exact numerical gradients relating the chosen objective function to the parameters that encode the permittivity structure of the training region. The algorithm then iteratively adjusts these parameters to maximize the overlap between the realized and desired modes of propagation \cite{minkov2020inverse, ceviche, Su2020Spins}. The result is a novel device geometry specialized to the user’s performance criteria.

Unfortunately, the set of physically realizable devices for inverse design methods is a small subset of the configuration space, leading to intensive investigation of devices which are amenable to these constraints such as photonic crystal-style devices \cite{burger2004inverse, miller2013photonic,minkov2020inverse, BorelTopologyWaveguide, LinPhotonicDirac} and metasurfaces \cite{ChristiansenTunableLens, PestourieMetasurfacesDemultiplex, ChungTunableMetasurface, MeemInverseFlatLens}. In general, these limitations require that certain restrictions be applied during the training process. For example, utilizing a continuous range of permittivities throughout the simulation domain greatly improves the ability of the algorithm to extremize the objective function, but continuity is not easily realizable with conventional manufacturing techniques and materials. To resolve this issue, a map from the parameterized design region to the set of actually manufacturable device configurations is used. These maps encode restrictions on the range and spatial distribution of the domain permittivity. A common technique in the literature is the use of nonlinear projection to design less versatile, but more easily manufacturable, binarized photonic devices \cite{piggott2015inversebinarized}. Such a device can be printed or created via lithography for uses in one-shot computing, but would be otherwise limited in general computing applications.

Rather than restricting the range of permittivities, we focus instead on limiting the spatial configuration by parameterizing the design region as a plasma metamaterial (PMM). Our PMM is a periodic array of cylindrical gaseous plasma elements (rods) which can be experimentally realized via the use of discharge lamps or laser-generated plasmas. PMMs are ideal candidates for inversely designed devices since the plasma density, and hence the permittivity, of each element within the PMM can be dynamically and precisely tuned through a continuous range of values ($\varepsilon\in(-\infty,1)$ for a collisonless plasma). This yields an infinite configuration space for training purposes \cite{configspaceppc} and allows a single device to serve multiple functions. Past work with inverse design electromagnetic devices achieved reconfigurability via methods such as refractive index changes \cite{ChristiansenTunableLens} in materials like liquid crystals \cite{ChungTunableMetasurface}. In our case, the relative permittivities of elements within the PMM device can be dynamically tuned by varying the gas discharge current. In addition, when the PMM is composed of elements that are small compared to the operating wavelength and the source is polarized properly, we can access localized surface plasmon (LSP) modes along the boundary between a positive permittivity background and a negative permittivity plasma element. Such surface modes can yield more complex and efficient power transfer \cite{tunableppc, 3dppc}, allowing for a particularly high degree of reconfigurability. Prior work by authors Wang and Cappelli with plasma photonic crystals (PPCs) has already highlighted the richness of this geometry in waveguide \cite{guideppc} and bandgap \cite{tunableppc, ppc, tunablepla} devices. The rich and varied physics of PMMs is well-matched to inverse design methods which promise to allow more holistic and efficient exploration of the configuration space.

\begin{figure}[htbp]
\centering
\includegraphics[width=\linewidth]{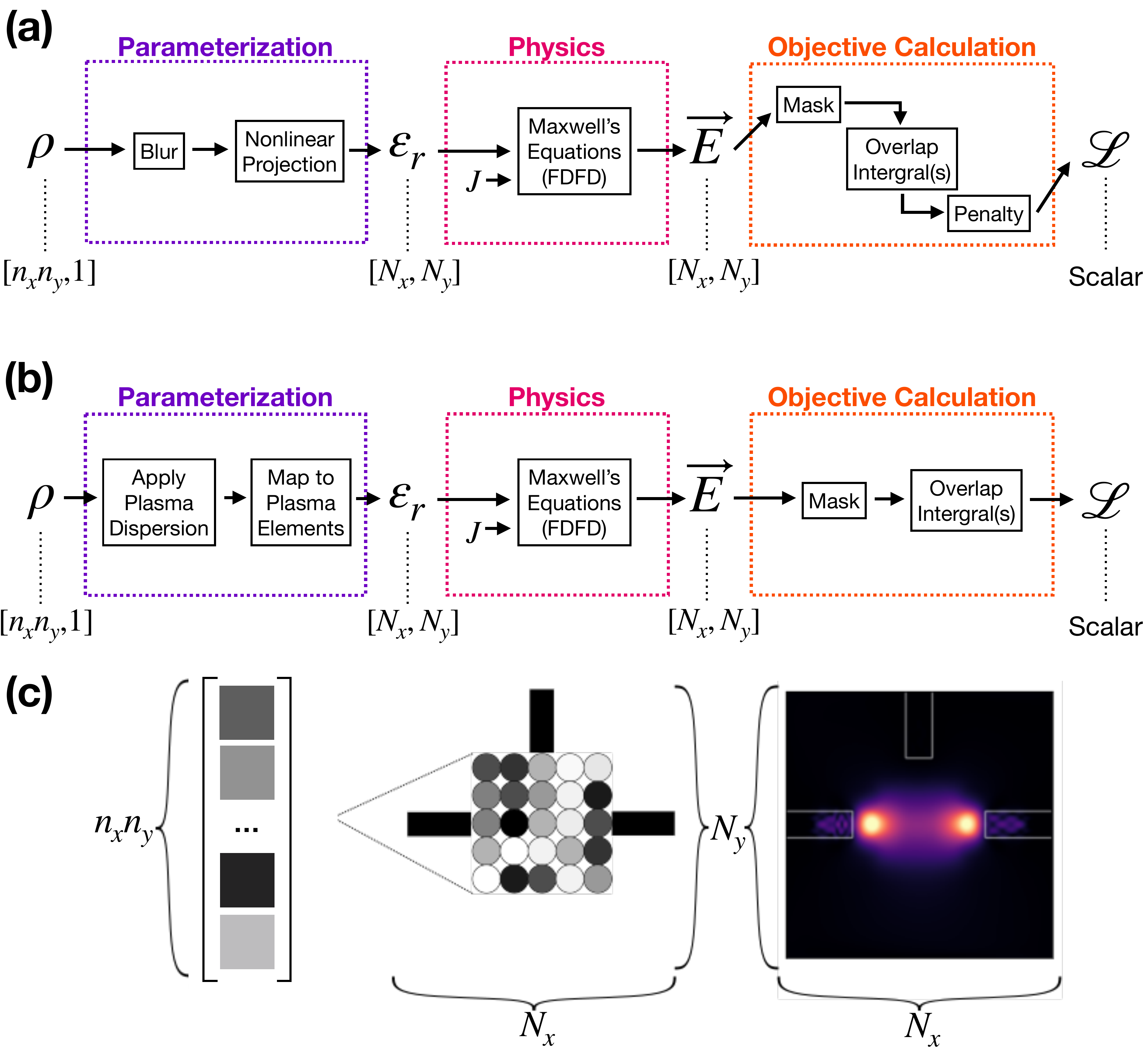}
\caption{Flow chart describing the algorithmic design of either (a) binarized material with penalized material use \cite{cevicheworkshop} or (b) array of plasma columns either in the TM mode ($\mathbf{E_x}$) or the the TE mode ($\mathbf{E_z}$). Though the parameterization stages are different, both configurations can be given identical sources and objective criteria. $J$ represents the modal source for the FDFD simulation. Examples of the training parameter vector $\rho$, permittivity matrix $\varepsilon_r$ and simulated $\textbf{E}$ fields for a simple PMM device are provided in (c).}
\label{fig:invdeschart}
\end{figure}

The design processes for binarized and PMM geometries are contrasted in Fig. \ref{fig:invdeschart}. Binarized geometries (Fig. \ref{fig:invdeschart}a) are challenging to optimize because of the requisite discontinuities. The jumps in permittivity from one material to another are difficult to handle in gradient-based methods that require the permittivity to vary continuously. Binarization can instead be approximated by any number of techniques, including gaussian blurring and highly nonlinear projections \cite{piggott2015inversebinarized}. In these configurations, the trainable parameters $\rho$ (of dimension $n_xn_y$) maintain a one-to-one correspondence with the individual pixels of the training region. In contrast, the number of training parameters in our PMM (Fig. \ref{fig:invdeschart}b) is determined by the dimensions of the plasma rod array, where there is a one-to-one correspondence between each plasma element and the $n_xn_y$ elements of $\rho$ (Fig. \ref{fig:invdeschart}c). In the end, both parameterization schemes result in a relative permittivity matrix $\varepsilon_r$ (of size $N_x\times N_y$) containing permittivity values at each pixel in the simulation domain. 

\section{Methods}
\begin{figure}[htbp]
\centering
\includegraphics[width=0.85\linewidth]{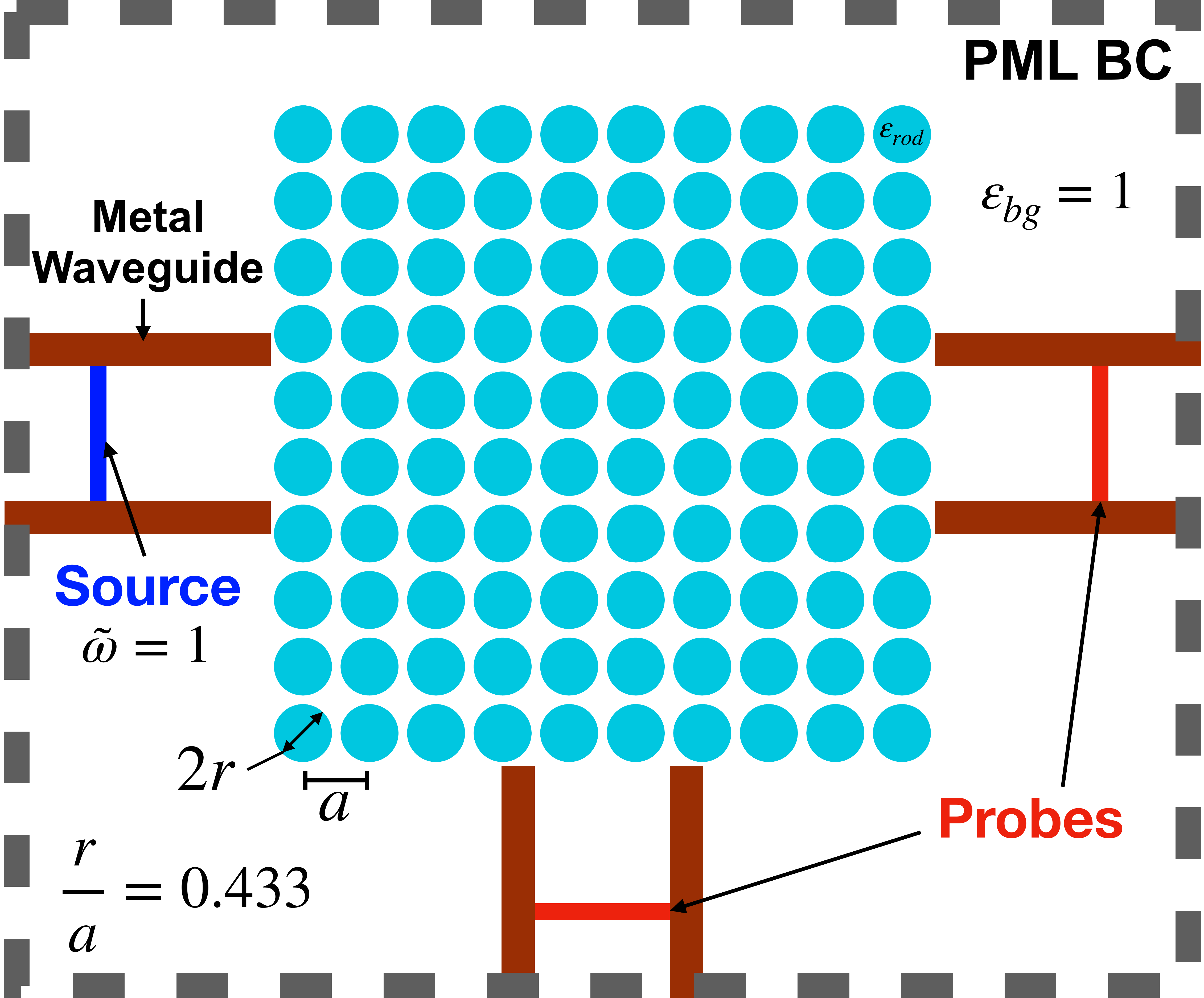}
\caption{Schematic diagram for the PMM waveguide device consisting of a $10\times10$ array of gaseous plasma rods suspended in air with metal waveguides functioning as the entrance and exit(s). The rod radius is chosen to be commensurate with current experimental facilities. The 'probe' slices represent where the mode overlap and intensity integrals that constitute the objective function are calculated. $\varepsilon_{bg}$ is the background permittivity. $\tilde{\omega}$ is the nondimensionalized operating frequency $\omega/(2\pi c/a)$.}
\label{fig:wvgschemchart}
\end{figure}

Fig. \ref{fig:wvgschemchart} provides a diagram of our PMM device in a directional waveguide configuration. Subsequent configurations exhibit the exact same array of tunable plasma elements but with different entrance/exit waveguides. The device is simply a 10$\times$10 array of tunable plasma rods suspended in air and spaced according to the limits of existing experimental facilites \cite{guideppc,ppc}. Modal sources are introduced at the input waveguide(s) and allowed to scatter through the training region before again collecting at the desired output waveguide(s). In this work, the fields are propagated through the domain via finite difference frequency domain (FDFD) simulations computed with Ceviche, an autograd-compliant electromagnetic simulation tool \cite{ceviche} that allows for calculation of the gradients of optimization objectives with respect to the input parameters that encode the permittivity domain. Note that the use of FDFD implies that all computed devices represent the steady state solution achieved after some characteristic time. The simulation domain was discretized using a resolution of 50 pixels per lattice constant $a$ and in each case presented $N_x=N_y=1000$ pixels. The only exceptions are the logic gate devices which were optimized at a resolution of 30 pixels$/a$ due to the high cost of the simulation. In those cases, the final device configurations were tested at the higher resolution of 50 pixels$/a$ and yielded results that were qualitatively identical. A perfectly-matched boundary layer (PML) $2a$ in width was applied along the domain boundaries. The polarization of the input source has a strong effect in devices of this nature, either \textbf{$E_z$} ($E$ out of the page), which we call the TM case, or \textbf{$E_x$} ($H$ out of the page), which we call the TE case. Previous work with PMMs indicates that both the TM and TE responses are tunable, with the latter benefiting from the presence of LSP modes \cite{righetti2018enhanced, tunableppc}, while the former makes more direct use of dispersive and refractive effects \cite{3dppc, guideppc, ppc}. The simulated $\mathbf{E}_x$ and $\mathbf{E}_z$ fields are masked to compute the field intensity and mode overlap integrals along planes of interest within the problem geometry. These integrals are used to calculate the objective $\mathcal{L}(\rho)$. Our simulation tool, Ceviche, is then used to compute numerical gradients of the objective with respect to the training parameters via forward-mode differentiation \cite{ceviche}. Ceviche uses the Adam optimization algorithm \cite{kingma2014adam} (gradient ascent, in essence) to iteratively adjust $\rho$ and thereby maximize $\mathcal{L}$. Optimization was conducted with learning rates ranging from $0.001-0.005$, and the default Adam hyperparameters $\beta_1=0.9$, and $\beta_2=0.999$.

In summary, our optimization problem is
\begin{align*}
    &\max_{\rho} \quad  \mathcal{L}(\mathbf{E}) \\
    &given \quad \nabla\times\frac{1}{\mu_0}\nabla\times \mathbf{E}-\omega^2\varepsilon(\omega,\rho)\mathbf{E}=-i\omega J.
\end{align*}
where $\mathcal{L}$ is the objective composed of a set of mode overlap integrals ($L_2$ inner product of simulated field with desired propagation mode), $\rho$ is an $n_xn_y$-dimensional vector that contains the permittivity values of each of the PMM elements, $\mu_0$ is the vacuum permeability, $\textbf{E}$ is the electric field, $J$ is a current density used to define a fundamental modal source at the input waveguide, $\omega$ is the field frequency, and $\varepsilon(\omega,\rho)$ is the spatially-dependent permittivity that is encoded by $\rho$. In practice, this permittivity distribution is achieved by varying the plasma density through control of the discharge current in each of the PMM elements according to the Drude model, which is where the dependence on $\omega$ arises. The non-dimensionalized plasma frequencies in $\rho$ are mapped to element permittivities via the Drude model with no collisionality (loss);
\begin{align*}
    \varepsilon=1-\frac{\omega_p^2}{\omega^2}
\end{align*}
where $\omega_p^2=\frac{n_ee^2}{\varepsilon_0m_e}$ is the plasma frequency squared, $n_e$ is the electron density, $e$ is the electron charge, $m_e$ is the electron mass, and $\varepsilon{_o}$ is the free-space permittivity. 

Of course, real gas discharge plasmas will be affected by some degree of collisional damping (represented by the loss parameter $\gamma$ in the Drude model). Such damping can be brought to very low levels in practice by reducing the discharge pressure, but this can come at the expense of plasma density (and correspondingly, plasma frequency), which affects the plasma dielectric constant. Furthermore, the neglect of collisional damping can lead to inaccurate results, particularly in the TE polarization where LSP modes can occur. Experimental noise/error in discharge current can also have an effect on device functionality; see Supplemental Material at [URL will be inserted once published] for a sensitivity study that was conducted on the optimal devices presented in the following sections. We choose to ignore collisions and other non-ideal factors in this particular study for several reasons. First, experimental results in our prior studies of PPCs where we operated primarily with TE polarized incident fields have agreed reasonably well with the collisionless Drude model \cite{guideppc, ppc}. In a study focused on a PPC bandgap device that utilized the LSP resonance, we showed experimentally that the resonance is present and strong for a relatively high collisionality where $\gamma/\omega_p=0.12$ \cite{righetti2018enhanced}. Second, the optimization algorithm used here is less stable and more time-consuming when the permittivity in the domain is of a complex data type, testing the limits of our available computational resources. Finally, this study is focused primarily on demonstrating that inverse design is readily applied to PMMs and its application represents a fruitful line of research. In future work, collisionality along with other non-idealities such as experimental noise and plasma non-uniformity within the discharges will be accounted for as we move closer to experimental realization.

\section{Results and Discussion}
\subsection{Directional waveguide}
\begin{figure}[htpb]
\centering
\includegraphics[width=\linewidth]{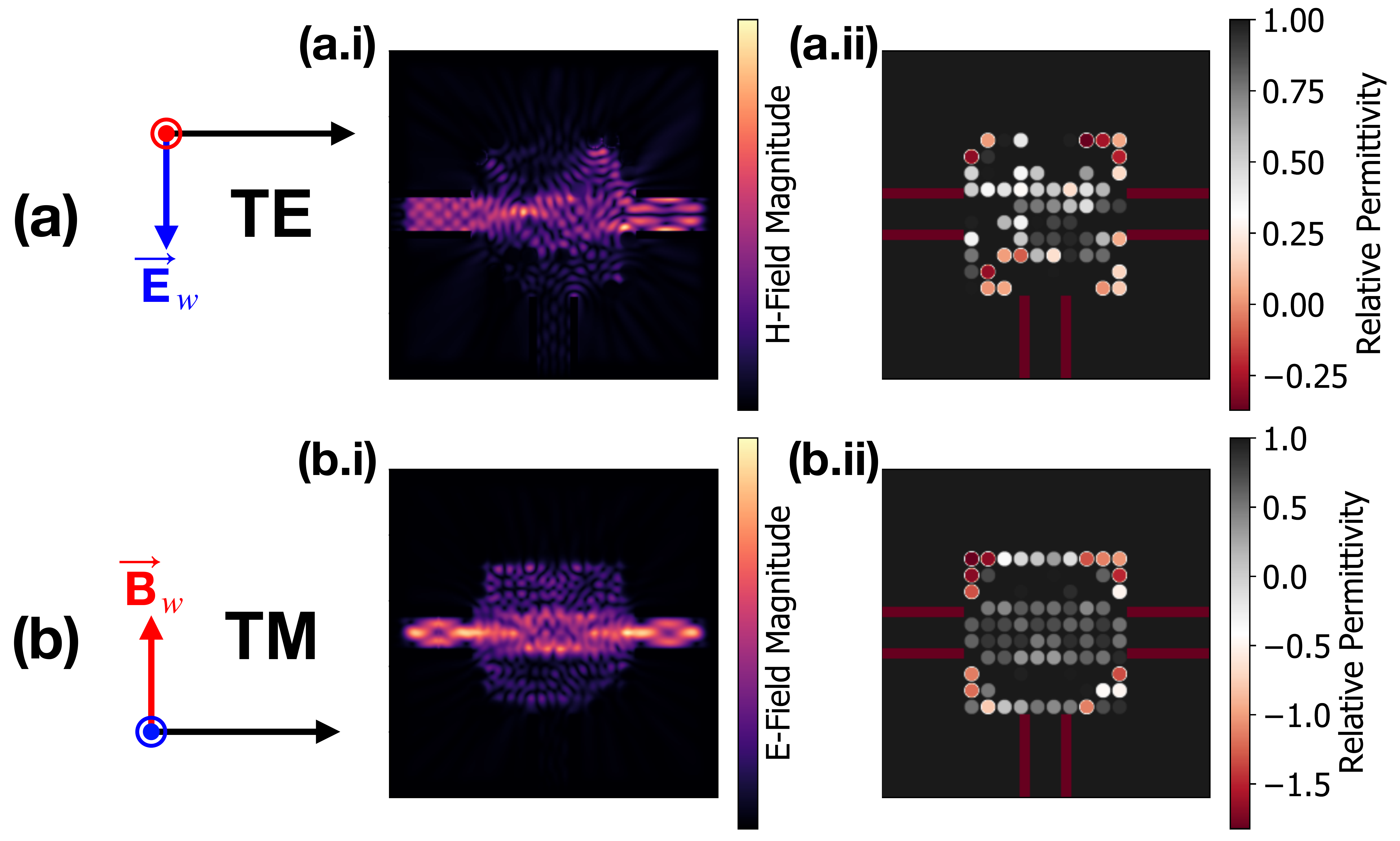}
\caption{(a) TE FDFD simulation showing (a.i) the field magnitude $|\mathbf{H_z}|^2$ ($\hat{\mathbf{z}}$ out of the page) for an optimized straight waveguide and (a.ii) the relative permittivity domain that gives rise to this behavior, where the maximum plasma frequency among the plasma elements is $\tilde{\omega}_p=1.17$. (b) FDFD simulation for the TM case showing (b.i) $|\mathbf{E_z}|^2$ and (b.ii) the relative permittivity with a maximum plasma frequency of $\tilde{\omega}_p=1.68$.}
\label{fig:straightguides}
\end{figure}

Electromagnetic waveguides are fairly common in inverse design devices either as the primary function or as important building blocks to more complicated devices \cite{BorelTopologyWaveguide, AndradeInverseGuide}. Presented first in Fig. \ref{fig:straightguides} are field simulations and permittivities of straight waveguide devices designed for either the TM or TE polarization. Operating at $\omega=1\times(2\pi c/a)$, or a non-dimensionalized source frequency $\tilde{\omega}=\omega/(2\pi c/a)=1$, these represent the simplest type of optimization problem for this configuration. In both polarization cases, the domain was initialized with $r/a=0.433$,  background permittivity $\varepsilon_{bg}=1$, and $\varepsilon_{rod}=0.75$ for every plasma element in the array. The plasma frequency of each of the rods was unconstrained during training ($\omega_p\in[0,\infty]$). The rod permittivities were initialized at 0.75 to allow the source to reach the exit waveguides in the first training epoch which consistently led to higher field intensities at the desired exit after training. In this and all other simulations, the permittivity of the input and output waveguide walls was set to $\varepsilon=-1000$ to serve as a lossless metal. The objective is simply the $L_2$ inner product of the simulated field with an $m=1$ propagation mode at the location of the probe in the desired exit waveguide,
\begin{equation*}
\mathcal{L}_{wvg}=\int E\cdot E_{m=1}^*dl_{\text{desired exit}}.   
\end{equation*}
No penalty was needed to achieve a decent standard of objective functionality. We can see from Fig. \ref{fig:straightguides} that both the TE and TM modes yield good performance, with a slightly better result in the TM mode straight waveguide. The optimized distribution of plasma densities (hence relative permittivities) in the plasma rods for the TM case is not surprising, representing a rectangular bridge of lower refractive index than its surroundings. However, the TE case settled on a distribution that was unintuitive. 

\begin{figure}[htpb]
\centering
\includegraphics[width=\linewidth]{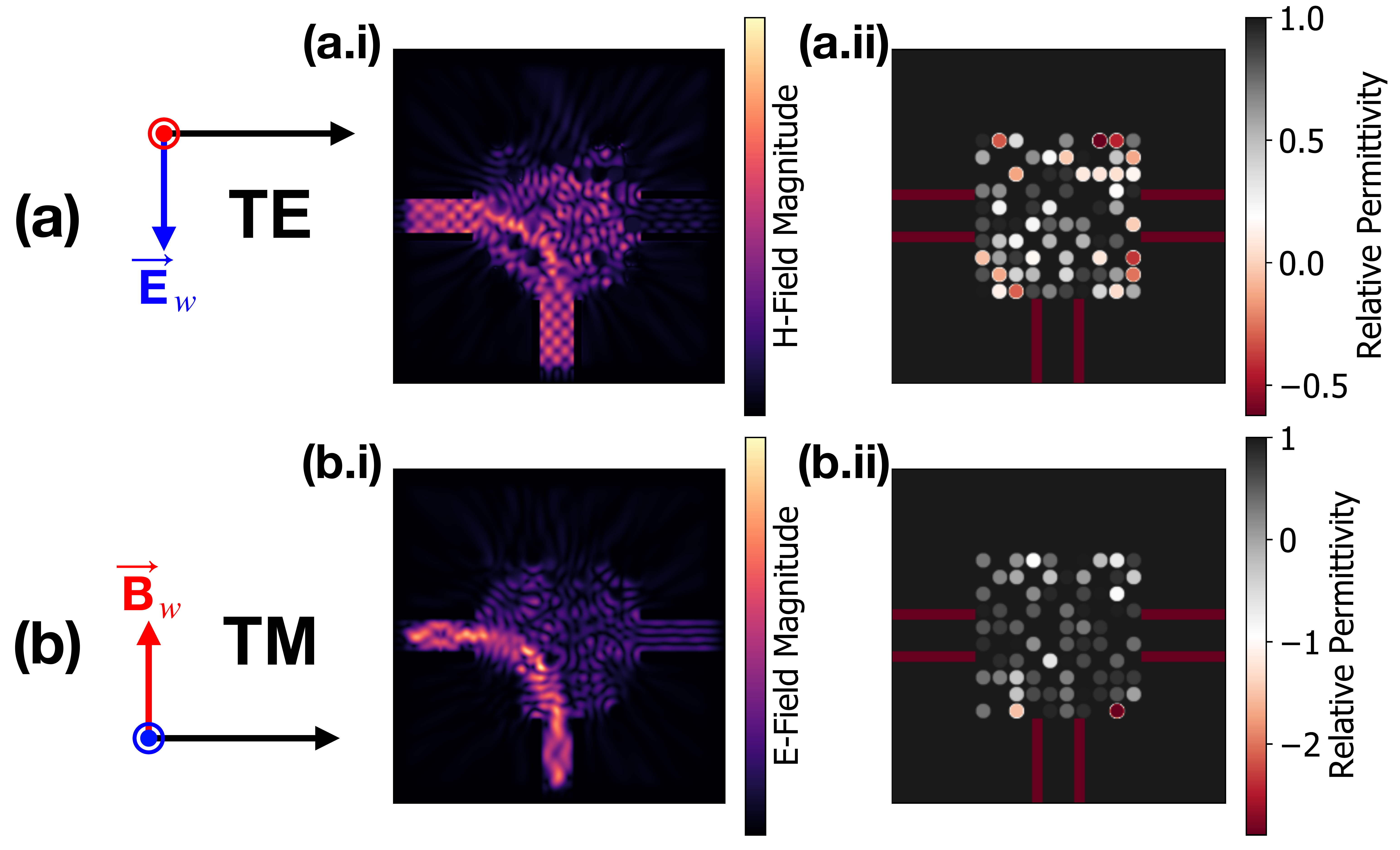}
\caption{(a) TE FDFD simulation showing (a.i) the field magnitude $|\mathbf{H_z}|^2$ ($\hat{\mathbf{z}}$ out of the page) for an optimized bent waveguide and (a.ii) the relative permittivity domain that gives rise to this behavior, where the maximum plasma frequency among the plasma elements is $\tilde{\omega}_p=1.27$. (b) FDFD simulation for the TM case showing (b.i) $|\mathbf{E_z}|^2$ and (b.ii) the relative permittivity with a maximum plasma frequency of $\tilde{\omega}_p=1.97$.}
\label{fig:bentguides}
\end{figure}

Next, we defined an objective intended to produce a waveguide with a 90-degree bend. The algorithm was initialized almost identically to the straight waveguide case, the only difference being that the objective overlap integral was computed at the bottom exit waveguide. No penalty was imposed at the incorrect exit waveguide as it was deemed unnecessary for satisfactory performance. The results for this case are found in Fig. \ref{fig:bentguides}. Again we see that, while both cases result in good device performance, the TM case again slightly outperforms the TE case which had some field leakage into to the undesired output waveguide.

\begin{figure}[h!]
\centering
\includegraphics[width=1\linewidth]{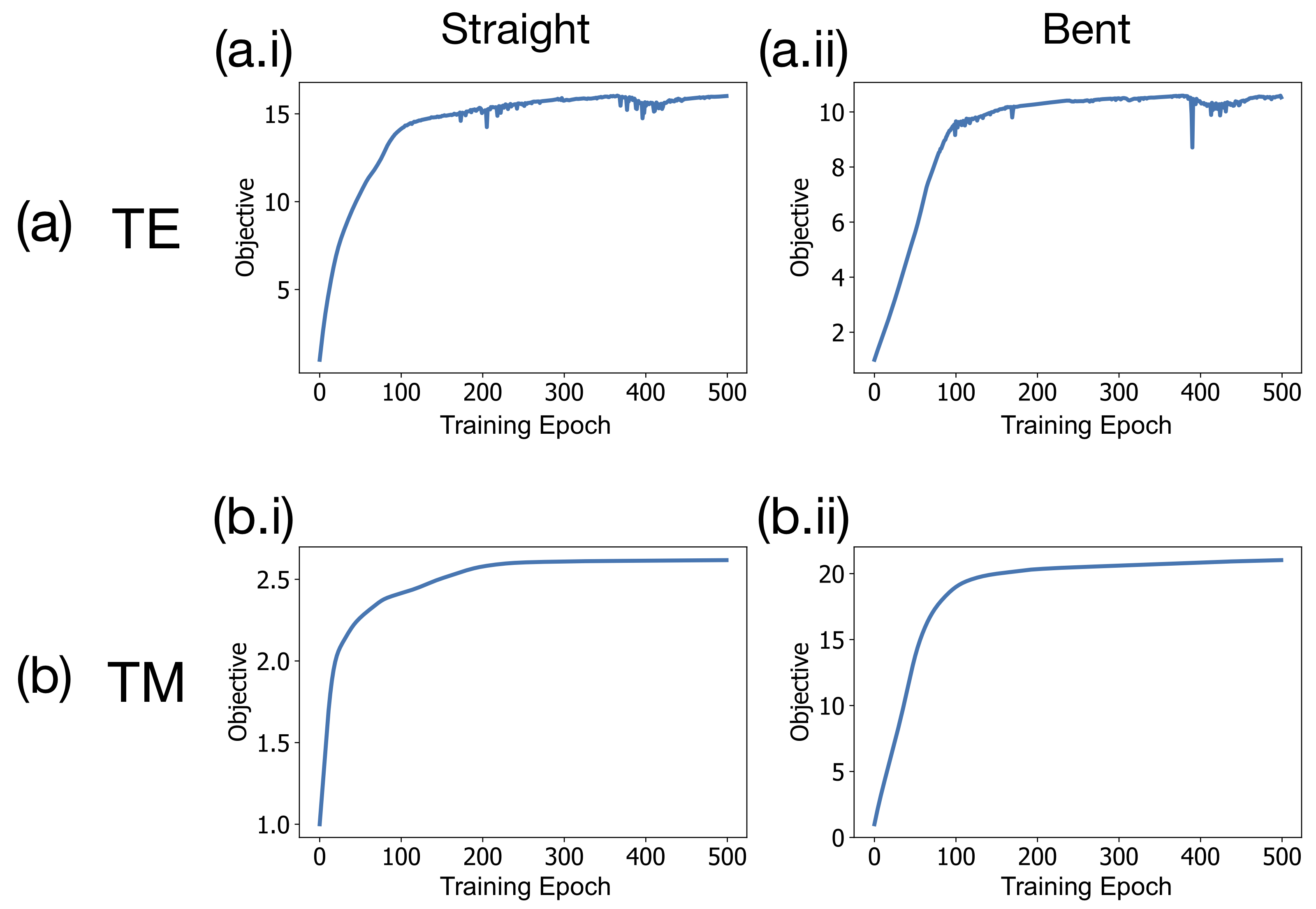}
\caption{Evolution of the objective functions for (a) the TE waveguides in the (a.i) straight waveguide case and (a.ii) the bent waveguide case. The same is presented for (b) the TM waveguides. The y-values of these plots are arbitrary as they are normalized by the initial (epoch 0) value.}
\label{guideobj}
\end{figure}

The devices presented in Figs. \ref{fig:straightguides} and \ref{fig:bentguides} appear to make expert use of the underlying electromagnetics. Inverse design allows a nuanced search of the configuration space, yielding more effective designs that preserve the input mode. Again, we emphasize that the final training results are not intuitive configurations. A human-reasoned design for this device may more closely resemble that of Wang et al. in ref. \cite{guideppc}, where the elements along the desired path of propagation are made equivalent to the background permittivity (i.e., the plasma is turned off) and the remainder of the rods are activated at the lowest permittivity (highest plasma frequency) possible to approximate a metallic waveguide. Instead, the algorithm settled on a configuration where a high contrast in permittivity of neighboring plasma columns is only present near the edges of the device to prevent leaking. The remaining elements are given a seemingly random distribution of plasma frequencies, keeping in line with typical results of inverse design schemes where strange structures often arise \cite{piggott2015inversebinarized, molesky2018inverse}. Fig. \ref{guideobj} graphs the evolution of the objective while training for these waveguide devices. The curves suggest that a local maximum was achieved in each optimization scheme. The spurious changes in the TE objectives are likely due to the excitation of LSPs. Because we use a collisionless plasma model, once the permittivity of the plasma elements is pushed to negative values by the parameter evolution, small changes in plasma frequency can excite localized surface plasmon resonances that strongly affect the overall wave propagation.

Overall, the results depicted in Figs. \ref{fig:straightguides} and \ref{fig:bentguides} are encouraging. Unlike binarized photonic devices which are static, the PMM geometry allows a single device to achieve both propagation modes because the permittivity of the device elements can be tuned electronically. Quickly switching between the straight and bent waveguide behavior, which is easily enabled by the controllable plasma current, opens up the possibility of transistor-like optical switches.

\subsection{Demultiplexer}
Next, we present a frequency demultiplexer designed to distinguish between frequencies of $\tilde{\omega}_1 =1$ and $\tilde{\omega}_2=1.1$. Inverse design methods have been used before to optimize wavelength demultiplexers in optical photonic structures \cite{piggott2015inversebinarized, PestourieMetasurfacesDemultiplex}. The optimization problem here is fundamentally different from the waveguiding case for three reasons: (1) a single device must now adhere to an objective based on two simulations, one for each operating frequency (2) the permittivity distribution is slightly different for the two frequencies because of the Drude dispersion relation and (3) a penalty is required to minimize leakage into the wrong output waveguide. The third difference is especially subtle since we want to discourage any leakage, not just a spurious $m=1$ mode. This means we cannot use the same mode integral as before. Instead we simply penalize the simulated field intensity at the incorrect exits: $\int |E|^2dl$. With all this considered, the demultiplexer objective is:
\begin{widetext}
\begin{equation*}
\mathcal{L}_{mp} = \left(\int E_{\omega1}\cdot E_{m=1}^*dl_{\omega1\text{ exit}}\right)\left(\int E_{\omega1.1}\cdot E_{m=1}^*dl_{\omega1.1\text{ exit}}\right)-\left(\int |E_{\omega1}|^2dl_{\omega1.1\text{ exit}}\right)\left(\int |E_{\omega1.1}|^2dl_{\omega1\text{ exit}}\right),
\end{equation*}
\end{widetext}
where $E_{\omega1}$ is the simulated field for the $\tilde{\omega}=1$ source and $E_{\omega1.1}$ is the simulated field for the $\tilde{\omega}=1.1$ source.

\begin{figure}[htbp]
\centering
\includegraphics[width=\linewidth]{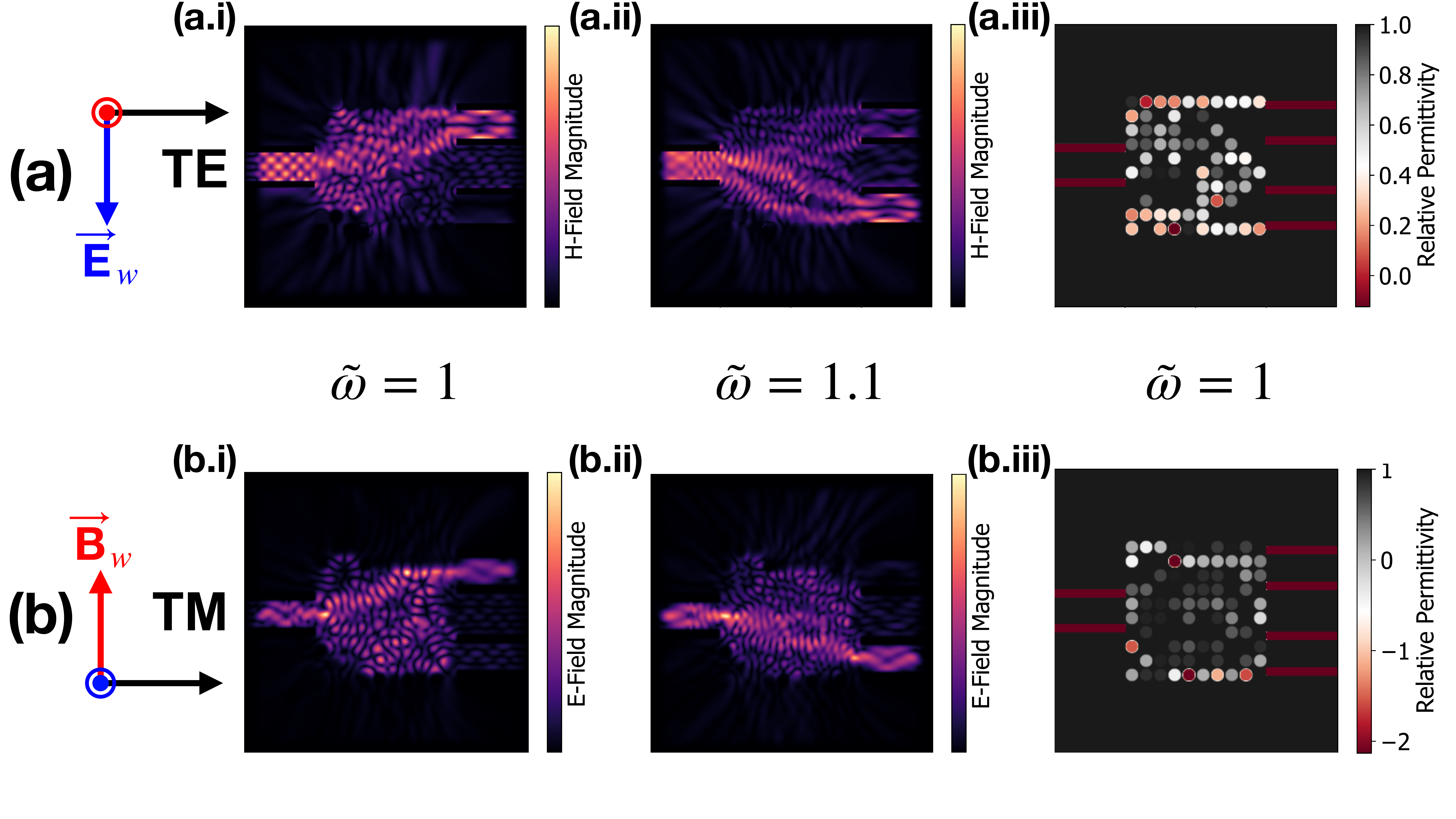}
\caption{(a) TE FDFD simulation showing (a.i) the field magnitude $|\mathbf{H_z}|^2$ ($\hat{\mathbf{z}}$ out of the page) for  $\tilde{\omega}{_1} =1$ and (a.ii)  $\tilde{\omega}{_1} =1.1$, along with (a.iii) the relative permittivity of the plasma rods when  $\tilde{\omega}{_1} =1$ where the maximum plasma frequency among the plasma elements is $\tilde{\omega}_p=1.17$. (b) FDFD simulations and  $\tilde{\omega}{_1} =1$ permittivities for the TM case where the maximum plasma frequency is $\tilde{\omega}_p=1.95$.}
\label{fig:multiplex}
\end{figure}

\begin{figure}[htpb!]
\centering
\includegraphics[width=1\linewidth]{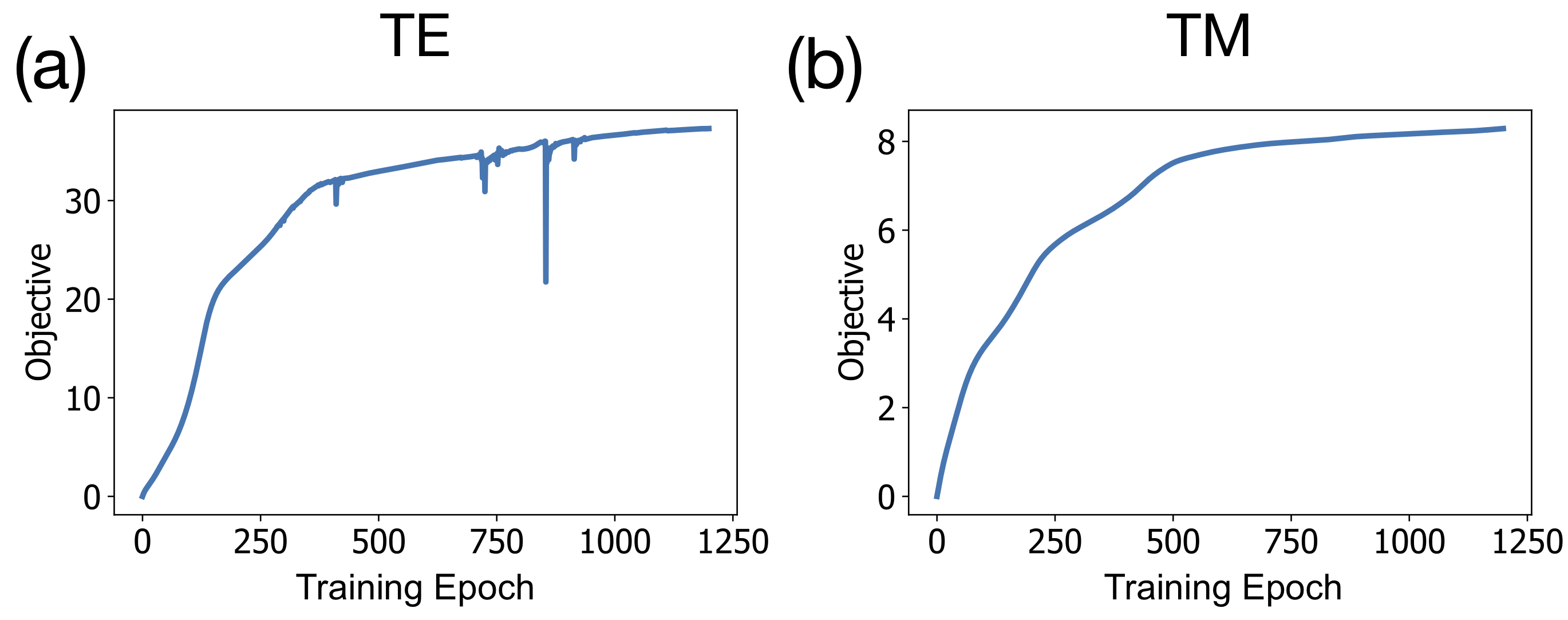}
\caption{Evolution of the objective functions for (a) the TE demultiplexer and (b) the TM demultiplexer. Once again, the y-axes on these plots are effectively arbitrary since they are normalized by the initial (epoch 0) value.}
\label{mpobj}
\end{figure}

After the 1250 epochs of training, the TE and TM devices presented in Fig. \ref{fig:multiplex} were obtained. Though these devices represent a jump in complexity, the objective functionalities were still accomplished quite well. The largest discrepancy is the minor leakage into the incorrect exit waveguide in the TE case operating at $\tilde{\omega}_1=1.1$. It is interesting to note that, despite the increased frequency sensitivity of the device in the TE mode when its constituent plasma elements have $\omega_p>\omega_{src}$ and therefore $\epsilon_{rod}<0$, the algorithm did not push rod permittivities deep into negative values. Perhaps bringing the source frequencies closer together might allow more effective usage of LSP excitations to differentiate the two frequencies. Fig. \ref{mpobj} shows the evolution of the objective for the two devices. It appears as before that a suitable local maximum is obtained in both cases. Once again, the final distribution of plasma frequency among the PMM elements for the demultiplexers provides little suggestion of their intended function.
  
\subsection{Low plasma frequency designs}
\begin{figure}[ht!]
	\centering
	\includegraphics[width=\linewidth]{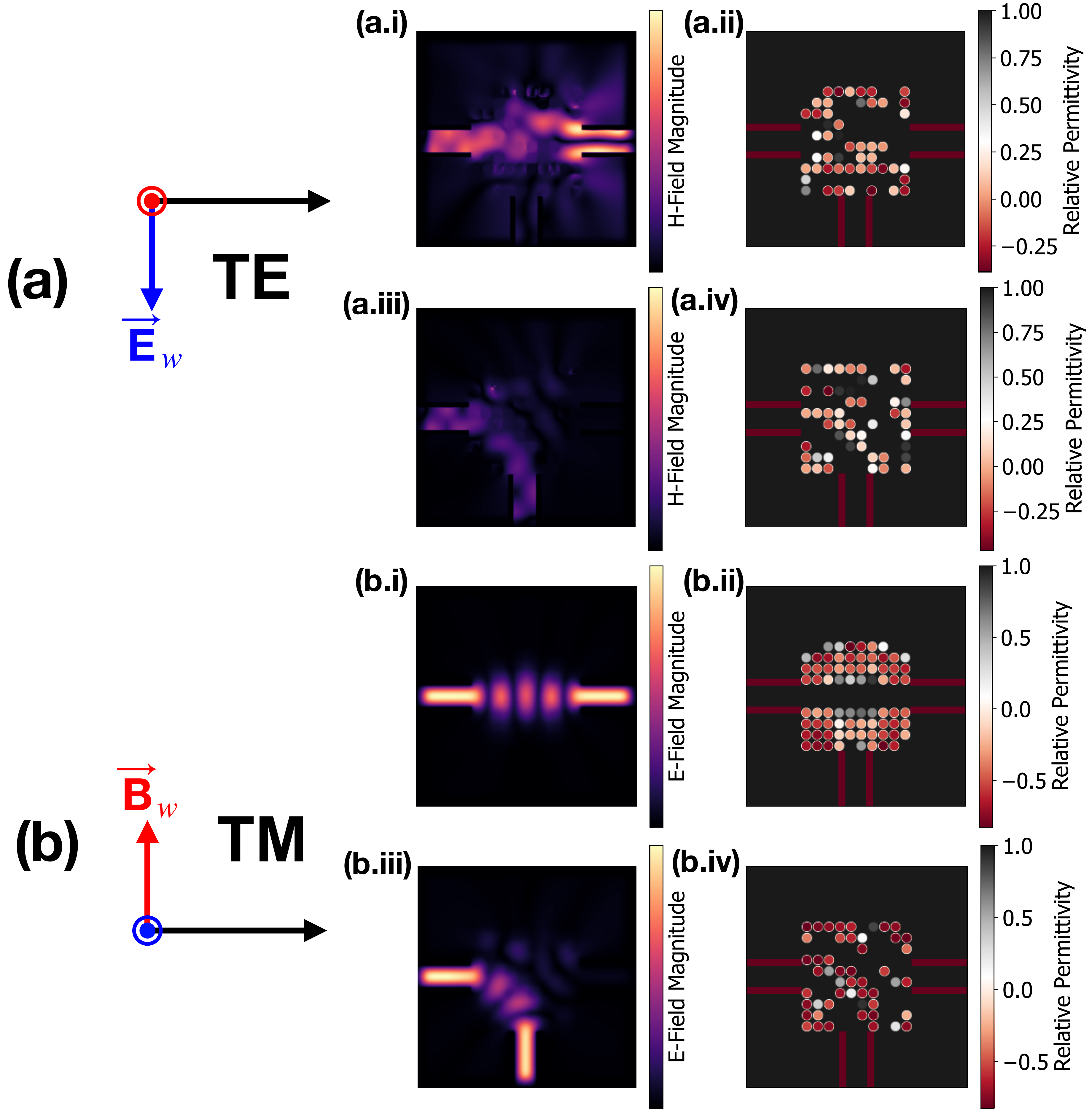}
	\caption{(a) TE FDFD simulation showing (a.i) the field magnitude $|\mathbf{H_z}|^2$ ($\hat{\mathbf{z}}$ out of the page) for an optimized straight waveguide and (a.ii) the relative permittivity domain that gives rise to this behavior, along with (a.iii) the field magnitude for the bent waveguide and (a.iv) its permittivity domain. (b) FDFD simulations and domains for the TM polarization. Among all of these cases, the highest plasma frequency is $\tilde{\omega}_p=0.338$ which corresponds to $\sim6.75$ GHz within existing experimental facilities.}
	\label{fig:LFguides}
\end{figure}

\begin{figure}[ht!]
\centering
\includegraphics[width=1\linewidth]{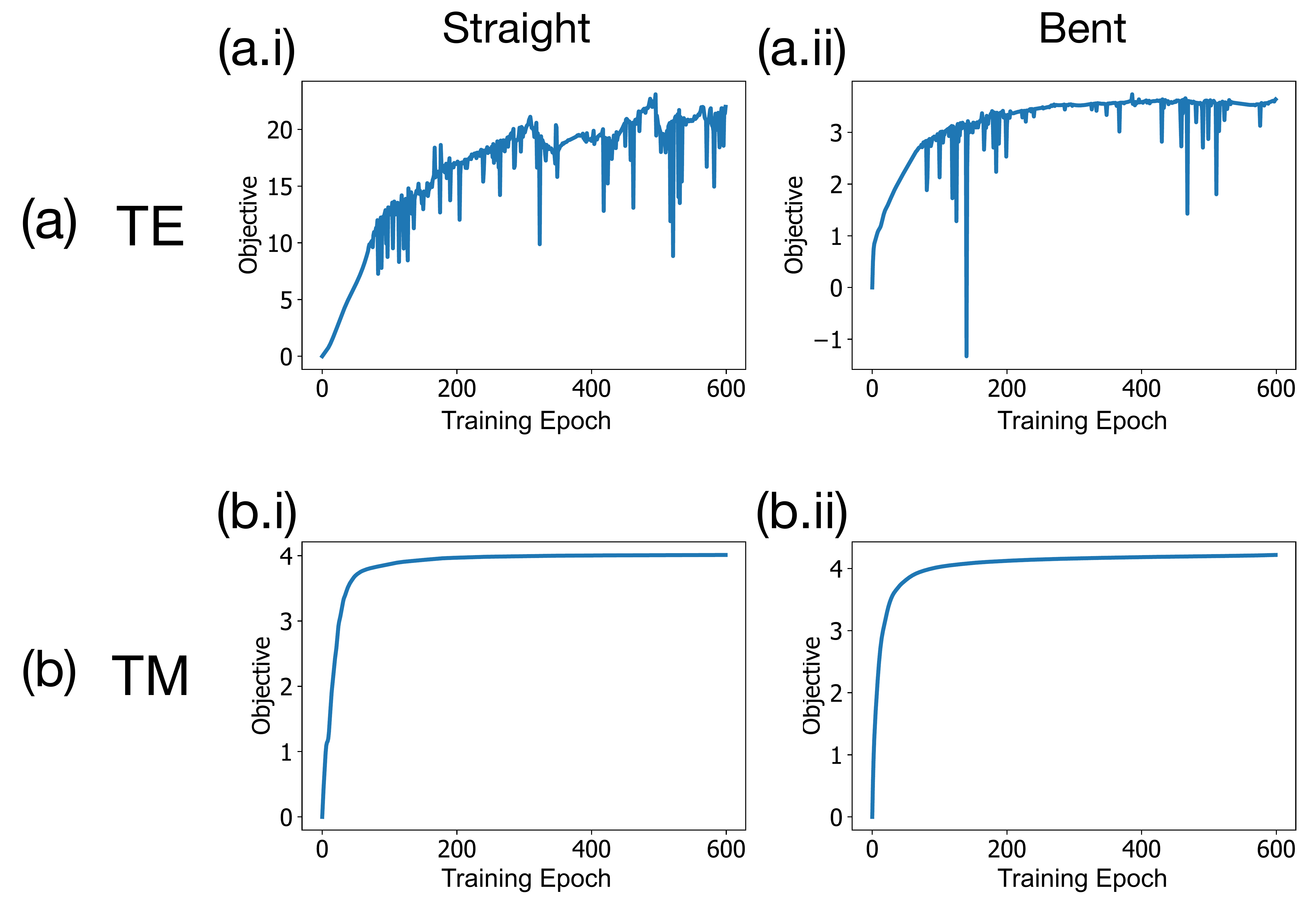}
\caption{Evolution of the objective functions for (a) the low-frequency TE waveguides in the (a.i) straight waveguide case and (a.ii) the bent waveguide case. The same is presented for (b) the TM waveguides. The y-axes on these plots are again arbitrary.}
\label{fig:LFGobj}
\end{figure}

Now, while these cases do exhibit compelling results, there is much to be said about their experimental realizability. Using the dimensions of plasma discharge tubes like those in refs. \cite{guideppc} and \cite{ppc}, the dimensionalized lattice frequency is estimated to be $\sim20$ GHz. Thus these devices call for plasma frequencies as high as $\sim39$ GHz, or, equivalently, plasma densities of $\sim2\times10^{19} m^{-3}$. These conditions are only possible through pulsed operation, which introduces complex transient behavior. In an effort to examine how these devices may be realized with currently existing plasma sources, we consider the case where the lowest frequency source allowed by the waveguides (width $2a$ yielding $\tilde{\omega} = 0.25$) is used and the plasma frequency is limited to $7$ GHz (an attainable quasi steady-state operation condition). These limits are enforced via a reparameterization of the optimization algorithm where an arctan barrier is employed. The results for these waveguide cases along with the evolution of the objective function are presented in Figs. \ref{fig:LFguides} and \ref{fig:LFGobj} respectively.

\begin{figure}[htbp]
\centering
\includegraphics[width=\linewidth]{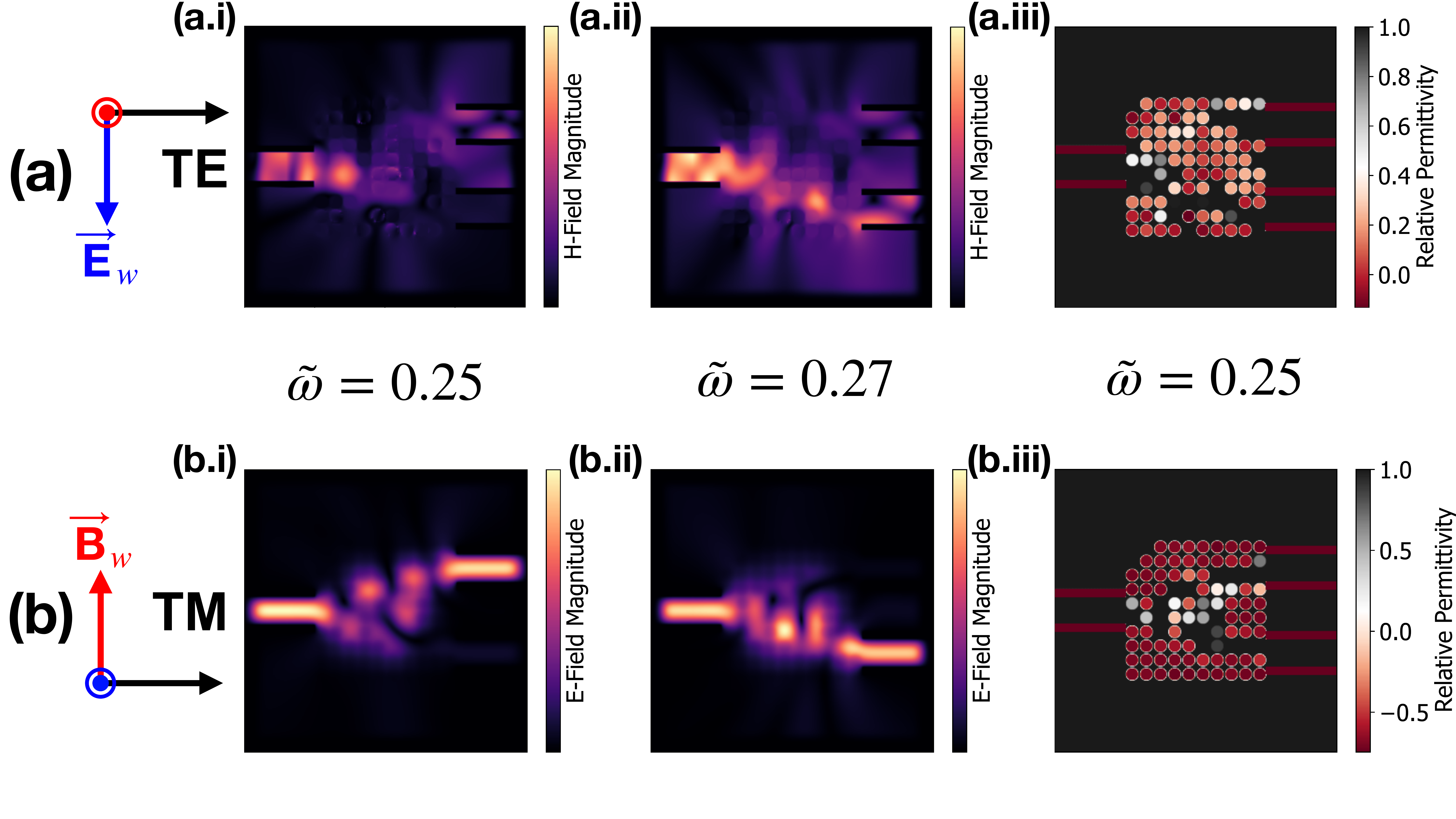}
\caption{(a) TE FDFD simulation showing (a.i) the field magnitude $|\mathbf{H_z}|^2$ ($\hat{\mathbf{z}}$ out of the page) for $\tilde{\omega}=0.25$ and (a.ii) $\tilde{\omega}=0.27$, along with (a.iii) the relative permittivity of the plasma rods when $\tilde{\omega}=0.25$ where the maximum plasma frequency among the plasma elements is $\tilde{\omega}_p=0.288$ which corresponds to $\sim5.75$ GHz within our existing experimental facilities. (b) FDFD simulations and $\tilde{\omega}=0.25$ permittivities for the TM case where the maximum plasma frequency is $\tilde{\omega}_p=0.358$ or $\sim7.13$ GHz experimentally.}
\label{fig:LFMultiplex}
\end{figure}

\begin{figure}[htpb]
\centering
\includegraphics[width=1\linewidth]{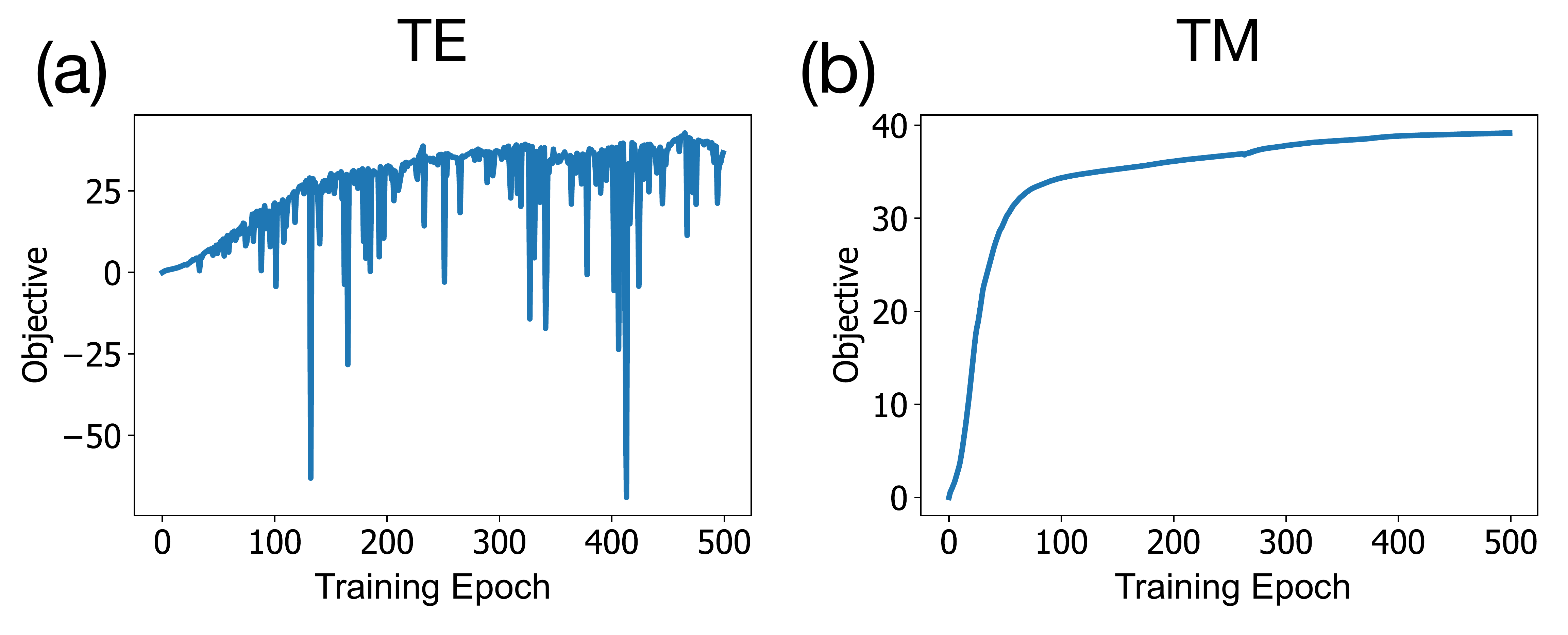}
\caption{Evolution of the objective functions in the low-frequency case for (a) the TE demultiplexer and (b) the TM demultiplexer. Once again, the y-axes on these plots are arbitrary.}
\label{fig:lfmpobj}
\end{figure}

We observe that for both objective functionalities, the TM-polarization case exhibits very strong performance while the TE case struggles to match its higher frequency alternatives. The reason for this is likely enhanced coupling to LSP modes. Since the plasma rods are now much smaller in radius than the wavelength of the source, LSPs are much more likely to appear and have a strong effect on the overall wave propagation. This is evident in the TE objective training curves which are far more erratic than their TM counterparts. Since the algorithm adjusts the permittivities of many rods at each step after calculating the gradients with respect to each parameter, the presence of LSP excitation can cause unexpected spurious losses that cause the objective to abruptly drop. In addition, the Adam optimization algorithm includes 'momentum' as part of its gradient ascent routine \cite{kingma2014adam}, so the objective is not guaranteed to increase monotonically. We see the same performance discrepancies in the demultiplexer, shown in Fig. \ref{fig:LFMultiplex}, and the demultiplexer objective, plotted in Fig. \ref{fig:lfmpobj}. Despite the losses in the TE polarization cases, these lower frequency simulations suggest that a reconfigurable device operating in the TM mode could be constructed with experimental discharge tubes such as those described in our prior studies \cite{guideppc,ppc}. More work needs to be done to identify objective functionalities that can reap the benefits of coupling into TE LSPs, perhaps with devices which require an extremely sensitive frequency response.

\subsection{Optical logic gates}
The success of these preliminary devices encourages experimentation with more complicated objectives and device geometries. Past work has established a connection between the wave dynamics inherent to electromagnetic phenomena and the complicated computations performed by recurrent neural networks (RNNs) \cite{nnanalog}. This type of analog computing appears to be the natural application of these PMM devices. The reconfigurability of gaseous plasmas means that the complicated classification tasks which are the staple of RNNs can be readily trained and implemented \textit{in-situ}. As a natural departure from traditional Von Neumann architectures, these optical computing platforms might yield significant advantages in bandwidth and throughput in an age where electronic computing appears to be reaching physical limits. Here we explore the application of inverse design to the fabrication of devices crucial to the development of more general optical computing platforms. Specifically, we consider the basic building block of traditional computing platforms: the electronic logic circuit. As physical realizations of boolean algebra, logic gates serve as the backbone for the core arithmetic, storage, and processing operations used in modern computing architectures. Optical logic gates could serve as the basic components of optical computing platforms \cite{DeMarinis2019}, and have been realized using photonic crystal systems \cite{FuOpticalGates, RaniAndGate, RaniPolIndepGate} as well as other techniques like semiconductor optical amplifiers \cite{BerrettiniGatesSOA}. We propose the application of inverse design PMM devices as a novel avenue for the fabrication of reconfigurable optical logic devices.

\begin{figure}[htpb]
	\centering
	\includegraphics[width=1\linewidth]{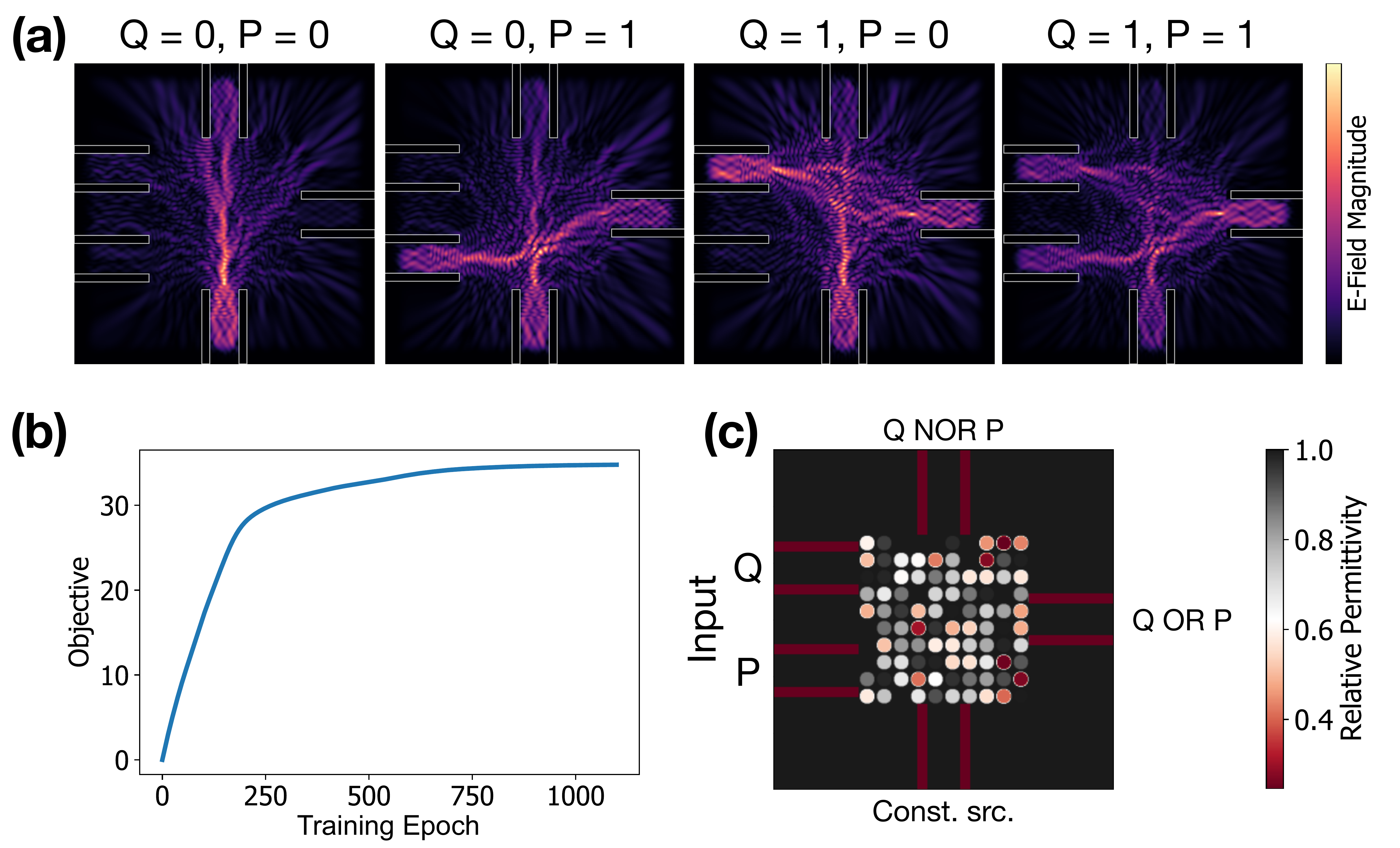}
	\caption{(a) E-field magnitudes for the OR gate in each of the input cases. (b) The evolution of the objective throughout training. (c) The relative permittivity domain along with labels for the device input and exit waveguides where the maximum plasma frequency is $\tilde{\omega}_p=0.869$.}
	\label{fig:or}
\end{figure}

\begin{figure}[htpb]
	\centering
	\includegraphics[width=1\linewidth]{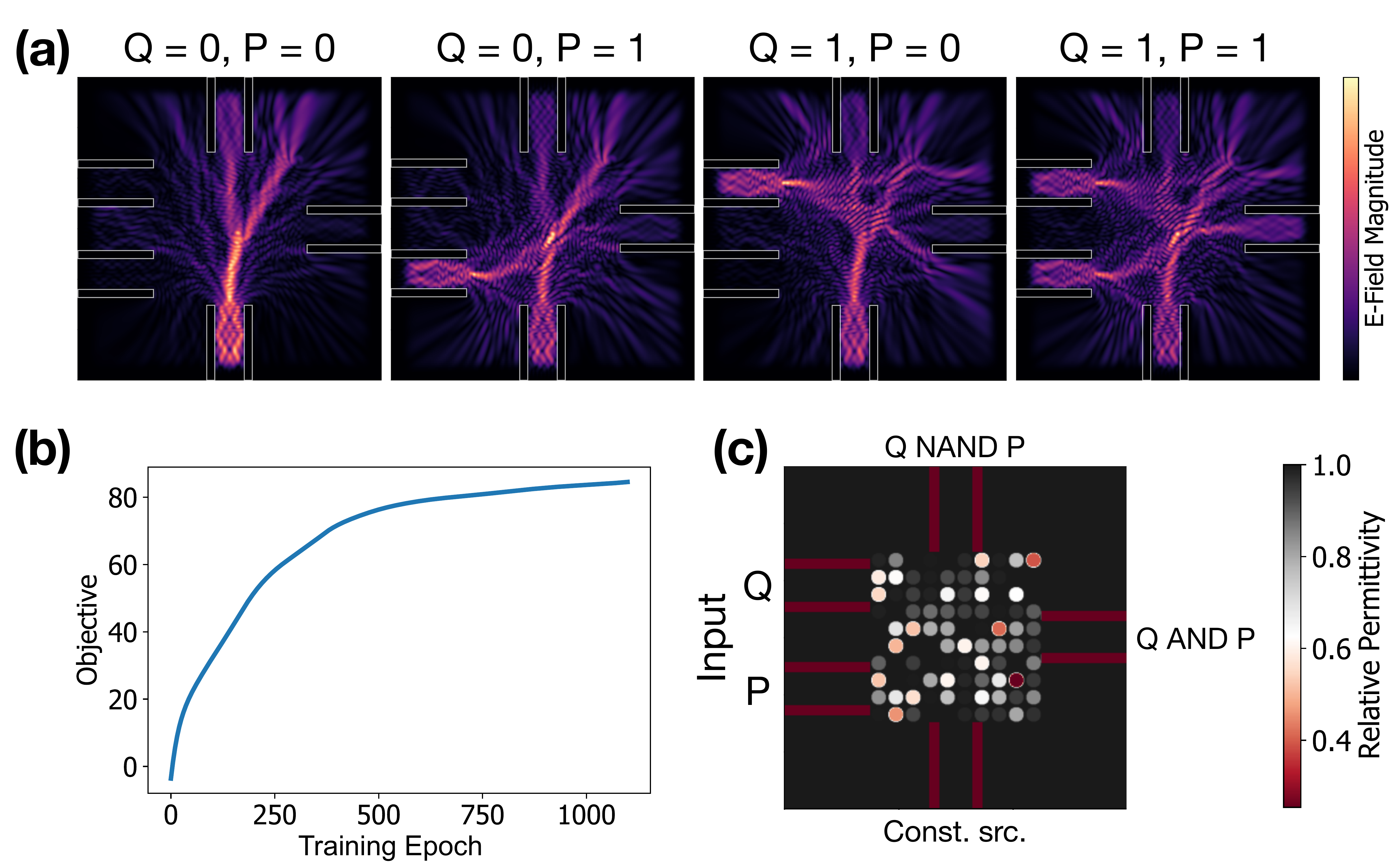}
	\caption{(a) E-field magnitudes for the AND gate in each of the input cases. (b) The evolution of the objective throughout training. (c) The relative permittivity domain along with labels for the device input and exit waveguides where the maximum plasma frequency is $\tilde{\omega}_p=0.864$.}
	\label{fig:and}
\end{figure}

These devices represent the most complicated objectives considered thus far, particularly since the boolean arithmetic is an inherently \textit{nonlinear} algebra. Because of this, we remove the plasma frequency constraints imposed for the simpler devices and use a TM polarized source frequency of $\tilde{\omega} = 2$ to allow for more complex propagation structures. As with the demultiplexers, the same device must be optimized to handle multiple cases depending on the source. Figs. \ref{fig:or} and \ref{fig:and} illustrate our platform for the realization of boolean logic through plasma arrays. Logical $1$s and $0$s are captured by the presence or absence of a modal source at each of the two input data waveguides. To account for situations where a logical $1$ is required while both databits are off (as is the case for a NOR or NAND signal), a constant field source is provided via the bottom waveguide. This in turn necessitates the addition of a "ground" sink in the case when a logical $0$ is required. The inclusion of this sink has the added benefit of dissuading reflection back into the input waveguides, which would hinder any efforts to link connectives and perform more complicated computations. 

While the previous cases were run entirely by the algorithm, the complex nature of this problem required the addition of many hyperparameters in the form of a set of variable weights ($w_1$-$w_8$ and $q_1$-$q_8$) that could be manually tuned after each optimization run. With additional computing power, one could design another shell of optimization around the Ceviche algorithm to obtain these weights automatically. The objectives used to train the permittivity domain for the gates were:
\begin{widetext}
\begin{align*}
    \mathcal{L}_{AND} &= w_1\int E_{Q0,P0}\cdot E_{m=1}^*dl_{\text{NAND}}-w_2\int |E_{Q0,P0}|^2dl_{\text{AND}}+w_3\int E_{Q1,P0}\cdot E_{m=1}^*dl_{\text{NAND}}-w_4\int |E_{Q1,P0}|^2dl_{\text{AND}}\\
    & +w_5\int E_{Q0,P1}\cdot E_{m=1}^*dl_{\text{NAND}}-w_6\int |E_{Q0,P1}|^2dl_{\text{AND}}+w_7\int E_{Q1,P1}\cdot E_{m=1}^*dl_{\text{AND}}-w_8\int |E_{Q1,P1}|^2dl_{\text{NAND}},\\
    \mathcal{L}_{OR} &= q_1\int E_{Q0,P0}\cdot E_{m=1}^*dl_{\text{ NOR}}-q_2\int |E_{Q0,P0}|^2dl_{\text{OR}}+q_3\int E_{Q1,P0}\cdot E_{m=1}^*dl_{\text{OR}}-q_4\int |E_{Q1,P0}|^2dl_{\text{NOR}}\\
    & +q_5\int E_{Q0,P1}\cdot E_{m=1}^*dl_{\text{OR}}-q_6\int |E_{Q0,P1}|^2dl_{\text{NOR}}+q_7\int E_{Q1,P1}\cdot E_{m=1}^*dl_{\text{OR}}-q_8\int |E_{Q1,P1}|^2dl_{\text{NOR}},
\end{align*}
\end{widetext}
where the weights are $(5,1,5,10,5,10,10.9,8.5)$ for the AND case and $(4,3,1,1.5,1,1.5,1,1)$ for the OR case. The results were found to be highly dependent on the choice of these weighting constants, suggesting that some form of automatic method for obtaining them might lead to better performance. In our case we manually tuned the weights following each run of optimization with Ceviche, where, for example, $w_8$ in $\mathcal{L}_{AND}$ would be increased following a run where the algorithm converged on a design with too much field intensity in the NAND exit for the $Q=1$, $P=1$ case. When using machine learning algorithms, it is common that hyperparameters such as the weights introduced here can have a major impact on the convergence of the algorithm to an effective design, and many methods exist for tuning hyperparameters toward that end \cite{bergstrahyperparams}. In our case, we did not have the computational resources to do so. Experimenting with different configurations of the objective determined that the simple linear combination of the field integrals above was the easiest to coax into good performance. Attempting to maximize the ratio between field intensities at the output and ground waveguides resulted in an optimization domain with many singularities, making convergence exceedingly difficult, and an objective consisting of a product of differences would yield the same objective for both the AND and OR gate since the parity of these objectives is the same. Nevertheless, these products and ratios of field integrals may be useful when employed in the optimization of the hyperparameter weights $w_i$ and $q_i$ should they be determined automatically as mentioned above.

The results shown in Figs. \ref{fig:or} and \ref{fig:and} are very compelling. In all cases, the field intensity at the correct output waveguide is substantially higher than at the incorrect output. Though some operating cases exhibit substantial loss, we understand that these losses are necessary when trying to achieve a nonlinear objective through linear dynamics. For example, the AND gate exhibits a large amount of leakage in the upper right corner of the device. This odd mode of propagation (halfway between the two exit guides) allows small changes in the sources to change which output waveguide sees a higher field intensity. Overall, these results show us that simple PMM device architectures can be reconfigured to achieve a wide array of highly non-trivial functions when inverse design is utilized.

\section{Conclusion}

There are several opportunities for further work based on the results discussed so far. The waveguide and demultiplexer devices designed here can be readily implemented and compared against simulated functionality using existing 2D PPC devices. Those parameters which are here taken as initial conditions  $(r/a, \varepsilon_{rod})$ can themselves be used as training parameters. The lattice constant in particular was noted to have a large impact on the trade-off between transmission to the output waveguide and control of propagation in the design region. Though in this study we chose the lattice constant to be commensurate with existing experimental equipment, more tightly-packed rods would likely lead to better performance. Consider also the prospect of a hexagonal or a 3D PMM arrangement. The non-dimensionalized operation frequency ($\omega$ in terms of $c/a$) may also prove to be an important optimization parameter as the dispersive physics which are the operational basis for these devices are highly frequency-dependent. In this study, aside from choosing the lowest frequency possible to reduce the plasma density requirements, other operating frequencies were determined without any substantial amount of reasoning. Of particular interest in the future are the effects of magnetization on the landscape of possible PMM devices. When used in conjunction with circularly polarized light, gaseous plasmas are capable of achieving $\varepsilon_r>1$, opening the door to even more complex logical operations. While several tools exist for the simulation and inverse design of photonic devices \cite{metanet}, more development is required on the application of inverse design to magnetized PMMs to incorporate such physics. Ceviche also includes a finite difference time domain solver that can be used to compute numerical gradients for optimization. Operating in the time domain would allow for a number of interesting designs to be pursued, such as analog recurrent neural networks which take advantage of nonlinearities in the plasma permittivity. The simulations in this study can also be further refined by adding collisionality (loss) and non-uniform plasma density profiles to the simulation. If the objective functionalities shown in this report can be achieved with these two non-ideal factors considered, then experimental realization would likely prove to be much more attainable. Finally, should experimental realization directly from simulation results prove difficult (we consider this to be a possible outcome), the device training could actually be carried out in-situ, with plasma element currents adjusted while the objective (transmission at a certain point, for example) is determined in real-time. In effect, the Ceviche forward-mode differentiation scheme that we use here could be carried out in the laboratory with real field measurements instead of simulated fields. This method would automatically take all non-ideal factors into account.

In conclusion, we consider the application of inverse design methods to the creation of highly optimized 2D PMM devices -- including waveguides, demultiplexers, and novel photonic logic gates -- for both TE and TM propagation modes. Objective functions of varying complexities are considered depending on the desired device. The optimization algorithm is shown to take advantage of the rich physics inherent to PMMs including tunable reflection, refraction, and the presence of localized surface plasmon modes. In the waveguiding and demultiplexer cases, simulated devices are shown with experimentally realizable parameters with few necessary changes to existing experimental infrastructure. However, further work is needed, both in simulations and experimental development, to enable device development for more complex objectives like boolean logic as well as other devices with complex wave propagation structures.

\begin{acknowledgments} 
A.A. would like to thank the Tomkat Center for support via the Energy Impact Fellowship. This research is also partially supported by the Air Force Office of Scientific Research through a Multi-University Research Initiative (MURI) with Dr. Mitat Birkan as Program Manager. J.A.R. acknowledges support by the U.S. Department of Energy, Office of Science, Office of Advanced Scientific Computing Research, Department of Energy Computational Science Graduate Fellowship under Award Number [DE-SC0019323].
\end{acknowledgments}

\bibliography{refs}

\end{document}